\documentstyle[aps,prl]{revtex}
\begin{document}
\draft
\flushbottom
\twocolumn[\hsize\textwidth\columnwidth\hsize\csname @twocolumnfalse\endcsname
\title{Polarized Wannier functions: {\it ab-initio} study
of the dielectric properties of silicon and gallium arsenide}

\author{Pablo Fern\'andez,$^{1,2}$ Andrea Dal Corso,$^{1,2}$
and Alfonso Baldereschi$^{1,2,3}$}

\address{$^1$ Institut de Physique Appliqu\'{e}e$-$Laboratoire
de Th\'eorie du
Solide (LTHS), \\
Ecole Polytechnique F\'{e}d\'{e}rale de Lausanne PHB-Ecublens, CH-1015,
Lausanne,
Switzerland \\
$^2$ Institut Romand de Recherche Num\'erique en Physique des Mat\'eriaux
(IRRMA), \\
PPH-Ecublens, CH-1015, Lausanne, Switzerland \\
$^{3}$ Istituto Nazionale per la Fisica della Materia (INFM), Italy.
}
\maketitle
\begin{abstract}
We present a first-principles calculation of the electronic properties
of crystalline silicon and gallium arsenide in a uniform electric
field.
Polarized Wannier-like functions which are confined in a finite region are
obtained by
minimizing a total-energy functional which depends explicitly
on the macroscopic polarization of the solid.
The polarization charge density and the electronic dielectric constant
are computed via finite differences.
The results coincide with those of the linear response approach in
the limit of vanishing electric field and infinite localization region.
\end{abstract}

\pacs{71.15.-m,71.15.Mb,77.22.-d}
]
\narrowtext

The study of materials in a finite electric field is
a difficult theoretical problem which is not definitively
settled yet~\cite{mo}.
The main difficulty is that the scalar potential which
describes the field is not periodic and is not
bounded from below: crystal momentum is no longer a good quantum
number and no ground state exists for the electrons of the solid.
{\it Ab-initio} studies of the response of crystals to uniform electric
fields have been done mainly with linear response theory (LRT).
Dielectric constants~\cite{LRT1}, piezoelectric tensors~\cite{LRT2},
second order non-linear susceptibilities~\cite{Dalcorso},
and Born effective charges~\cite{LRT1} of several
materials have been computed with density functional perturbation
theory (DFPT)~\cite{LRT1,Gonze}.
In practice however with these methods only the response to
{\it infinitesimal} electric fields can be studied.

A major issue is therefore the study of solids in a {\it finite} electric
field. Linear as well as nonlinear
susceptibilities could be extracted simultaneously from
such calculations, together with
important informations on technologically
interesting phenomena, such as electromigration,
which have eluded, until now, a detailed {\it ab-initio}
investigation~\cite{Kaxiras}.
Only a few direct calculations with finite
electric fields have been done with large supercells and
artificially periodic fields~\cite{diref,bernardini}.
Recently Nunes and Vanderbilt~\cite{Nunes} have proposed an approach
to circumvent the difficulties associated with finite electric
fields:
they have shown that localized Wannier functions (LWFs) allow one
to write a functional for the energy of a solid in a
uniform electric field. The ground state of this functional corresponds to
a meta-stable state for the real solid~\cite{nenciu},
but, in the limit of vanishing electric field,
the derivatives of the functional
provide the same dielectric properties
(polarization, linear~\cite{Nunes} and non-linear~\cite{Dalcorso}
dielectric constants) as the perturbative
approach. The method has so far been applied only to a simple
one-dimensional tight-binding model. In this Letter we demonstrate its
applicability within a self-consistent scheme by
computing the dielectric properties of silicon and gallium arsenide.
This study is of relevance also for `order N'
electronic structure calculations~\cite{mauri} since
LWFs are a key-ingredient for many of these methods.

In a previous work~\cite{fdmb}, we computed
LWFs of silicon and gallium arsenide in zero electric
field within a self-consistent scheme in the local density
approximation (LDA).
In this work, we consider these two materials in a finite electric
field and obtain their {\it polarized} WFs. We study
the ability of the latter to predict the electronic properties
of insulators in a uniform electric field and discuss the convergence of the
dielectric properties with increasing size of the localization
region (LR).

As underlined by Gonze, Ghosez, and Godby~\cite{ggg}
the validity of the Hohenberg and Kohn theorem~\cite{hk} for an infinite
insulating crystal in a macroscopic  electric field is questionable
and in principle one should add the macroscopic polarization as a basic
variable.
Martin and Ortiz~\cite{mo} have recently reformulated the DFT for
insulating crystals to include electric field effects.
However, no realistic polarization-dependent functional
has been proposed yet and in this work we remain at the standard
LDA level.
We use equations similar to those of Ref.~\onlinecite{mo} except for
the ``exchange and correlation field'', which stems from the
polarization dependence of the exchange and correlation energy.

We describe first how to implement
finite electric fields in our scheme~\cite{fdmb} based on LWFs.
We then show that the variation of the charge density induced,
at first order,
by a finite electric field, and computed with polarized LWFs,
converges to the values provided by LRT.
We also analyze the induced macroscopic polarization and show that,
increasing the size of the LRs,
its derivative with respect to the field converges exponentially.
The asymptotic values, for vanishing field, are close to the DFPT results.
Finally we present examples of polarized WFs.

The electronic ground state of a periodic crystal containing $N$
interacting electrons can be described either by
$N/2$ independent Bloch orbitals or by $N/2$ WFs.
One way to obtain approximate WFs for solids has been presented
in Ref.~\onlinecite{mauri} and implemented in a self-consistent DFT scheme
in our previous work~\cite{fdmb}.
The method is based on the minimization of a total energy functional
$E_{tot} \left[ \{ | v_{0,n} \rangle \} \right]$ which includes implicitly
orthogonalization constraints with respect to $M = N_e/2$ localized
orbitals $\{ | v_{0,n} \rangle \}$, where $N_e$ is the number of
electrons per unit cell, $n = 1, \ldots, M$ is the band index,
and the
subscript $0$ indicates that the orbital is centered in the unit cell
containing the origin.
The periodicity of the crystal is exploited by requiring that the remaining
orbitals are obtained by translating those of
the first cell by all the Bravais lattice vectors
${\bf R}_l$: $|  v_{l,n} \rangle =
\hat{T}_{ {\bf R}_l } | v_{0,n} \rangle$.

As already mentioned, the Hamiltonian of an infinite, periodic crystal
in presence of a uniform static electric field
${\bf E}$~\cite{notaE},
is not bounded from below. However
if the degrees of freedom
$ \{ | v^{{\bf E}}_{0,n} \rangle \}$ are constrained to be localized
in finite regions and the electric field
is sufficiently small, the functional~\cite{Nunes}
\begin{equation}
 F\left[ \{ | v^{{\bf E}}_{0,n} \rangle \} \right] =
 E_{tot} \left[ \{ | v^{{\bf E}}_{0,n} \rangle \} \right]
 - \Omega {\bf P} \cdot
 {\bf E}
 \label{deffunct},
\end{equation}
(where $\Omega$ is the unit-cell volume and the macroscopic
electric polarization ${\bf P}$ is the sum of the
electronic ${\bf P}_e$ and ionic contributions)
has a well defined minimum which represents a meta-stable state
for the real solid~\cite{Nunes,nenciu}.
In the limit where the size of the LRs becomes infinite,
the maximum allowed field goes to zero.
For any non-vanishing value of the electric field ${\bf E}$,
the second term in Eq.~(\ref{deffunct}) will cause a change in the periodic
density and also a variation of the macroscopic polarization.
In our calculations, the value of the
screened electric field ${\bf E}$ is kept fixed, while the macroscopic
polarization changes self-consistently. Physically we compute variations
of the charge density,
of the macroscopic polarization, and of the total energy
with respect to the screened uniform electric field
instead of the bare external electric field ${\bf E}_0$.

At zero field, the functional is minimized by
almost orthogonal orbitals which, for sufficiently large LRs,
are a good approximation to a set of WFs $\{ | w_{0,n} \rangle \}$
of the solid. Similar sets of WFs can also be obtained  through
a unitary transformation of the Bloch orbitals
with an appropriate choice of the phases~\cite{marzari}.
In presence of a finite field, the functions
$ \{ | v^{{\bf E}}_{0,n} \rangle \}$ which minimize the functional are
still approximately orthogonal and, for large LRs, approximate a set of {\it
polarized}
WFs for the solid.
The electronic contribution ${\bf P}_e$ to the macroscopic polarization ${\bf
P}$
can be related~\cite{pol} to the
{\em centers} ${\bf r}_n = \langle w_{0,n} | {\bf r} | w_{0,n} \rangle$ of the
WFs of the system by
$\Omega {\bf P}_e = -2 e \sum_{n=1}^{M} {\bf r}_n$.
This definition can be {\em extended} to non-zero values of the
screened macroscopic electric field and to
LWFs by~\cite{Nunes}:
\begin{equation}
\Omega {\bf P}_e\left[ \{ | v^{{\bf E}}_{0,n} \rangle \}, {\bf E} \right]
 = -2 e  \sum_{n=1}^{M} {\bf r}_n({\bf E}), \label{polE}
\end{equation}
with ${\bf r}_n({\bf E})$ given by
\begin{eqnarray}
 {\bf r}_n({\bf E}) = \sum_{l} \sum_{m=1}^{M}
                      \langle v^{{\bf E}}_{0,n} | {\bf r}
                      | v^{{\bf E}}_{l,m} \rangle Q^{l,0}_{m,n}({\bf E}),
\label{centerE}
\end{eqnarray}
where $e$ is the electronic charge, and
$Q^{l,0}_{m,n}({\bf E}) = 2 \delta_{l,0} \delta_{m,n} -
\langle v^{{\bf E}}_{l,m} | v^{{\bf E}}_{0,n} \rangle$ is the first-order
approximation to the inverse of the overlap matrix between
LWFs.

This extended definition of ${\bf P}_e$ allows one to study the dielectric
properties of insulating crystals in finite electric fields.
The validity of this definition rests on the agreement of the values of
the dielectric properties computed in the limit of vanishing field
with the LRT results~\cite{Nunes,Dalcorso}. In this limit the equivalence
between the two methods can be demonstrated analytically.
The electronic dielectric tensor
$\epsilon^\infty_{\alpha \beta} ({\bf E})$ is related to
the macroscopic electronic polarization ${\bf P}_e$ by
\begin{equation}
 \epsilon^\infty_{\alpha \beta}({\bf E}) = \delta_{\alpha \beta}
  + 4 \pi \frac{d P_{e, \alpha} ({\bf E})}{d E^\beta}
 \label{epsi}
\end{equation}
where $\alpha$ and $\beta$ indicate  Cartesian coordinates.
It  can be obtained by finite differences of the
macroscopic polarization ${\bf P}_e({\bf E})$ for different values of
the screened electric field ${\bf E}$. ${\bf P}_e({\bf E})$  refers
to
${\bf P}_e\left[ \{ | v^{{\bf E}}_{0,n} \rangle \}, {\bf E} \right]$ of
Eq.~(\ref{polE})
computed for the orbitals $\{ | v^{{\bf E}}_{0,n} \rangle \}$
which correspond, for a given electric field ${\bf E}$, to the
minimum of the functional $F$ .

The search for the minimum of the total-energy functional $F$ requires
its functional derivative with respect to the variational
degrees of freedom $\langle {\bf r} | v^{{\bf E}}_{0,n} \rangle$.
Differentiating Eq.~(\ref{deffunct}) with respect to
$\langle {\bf r} | v^{{\bf E}}_{0,n} \rangle$, the polarization term gives
an additional contribution with respect to the zero-field case,
which can be recast as:
\begin{equation}
 4 \sum_{l,m} \left[ \hat{T}_{{\bf R}_l} ( {\bf r} | v^{{\bf E}}_{0,m} \rangle
)
 Q^{l,0}_{m,n} - | v^{{\bf E}}_{l,m} \rangle \langle v^{{\bf E}}_{l,m} | {\bf
r} | v^{{\bf E}}_{0,n} \rangle \right]
\cdot e {\bf E}. \label{supterm}
\end{equation}

The  LWFs have been represented on a uniform
real-space mesh and $\langle {\bf r}| v^{{\bf E}}_{0,n} \rangle$
is allowed to be non-zero only inside cubic regions
with size $2 a_{LR}$.
As in the zero-field case, the localization of the
LWFs makes the
summations in Eqs.~(\ref{centerE}) and (\ref{supterm}) finite.
The periodic
valence charge density $n({\bf r};{\bf E})$
(whose expression is the same as in the zero-field case)
is a sum of localized contributions. The Fourier components
$\tilde{n}({\bf G};{\bf E})$ allow us
to compute the Coulomb electrostatic energy per unit cell as
$E_{H} = \frac{\Omega}{2} \sum_{{\bf G} \neq {\bf 0}} \tilde{n}^\star
({\bf G};{\bf E}) \tilde{V}_H({\bf G}; {\bf E})$,
where $\tilde{V}_H({\bf G}; {\bf E})$ is the Fourier transform of the Hartree
potential.
The ${\bf G}=0$ component of the Hartree potential is included in the
screened electric field ${\bf E}$ and it is not explicitly evaluated.

We have applied this approach to crystalline silicon and gallium arsenide.
The unit cell contains
$2$ atoms and the electronic ground-state can
be described by $M=4$ LWFs which
are confined within LRs centered on the four bonds in the unit cell at the
origin.
Our calculations have been carried out within DFT-LDA~\cite{ldanota}
and with norm-conserving
pseudopotentials in the Kleinman-Bylander form~\cite{kb}.
All technical details and parameters are the same as those given in
Ref.~\onlinecite{fdmb}.

In Fig.~\ref{fig1}, we show the charge density induced by a field in the
$[100]$ direction
and with an intensity equal to $e \left| {\bf E} \right| = 10^{-3}$ a.u.
$\approx
5.14 \times 10^{8}$ V/m.
The induced valence charge density $\Delta n({\bf r})
= n({\bf r}; {\bf E}) - n({\bf r}, {\bf 0})$
computed with LWFs obtained for different values of $a_{LR}$
is compared in Fig.~\ref{fig1}(a) with the result of the conventional linear
response
approach~\cite{LRT1}
$n^{(1)}({\bf r}) = {\bf E} \left.
\frac{\partial n({\bf r})}{\partial {\bf E}} \right|_{{\bf E}=0} $
obtained using a Bloch orbital representation
of the electronic wave-functions~\cite{nota2}.
The weak intensity of the field
ensures a linear behaviour for both silicon and gallium arsenide.
Fig.~\ref{fig1}(a) shows that all the features of the charge density linearly
induced by the field are correctly reproduced with small LRs.
Localization regions containing $216$ atoms ($a_{LR}/a = 34/24$) are large
enough to obtain an induced charge density which is practically
undistinguishable
from the LRT result on the scale of the figure. The difference
is everywhere less than $4.7\times 10^{-3}$ e/cell for silicon and
$6.1\times 10^{-3}$ e/cell for gallium arsenide. In the case of
gallium arsenide, this error is smaller
than that associated with the use of a semi-local pseudopotential
instead of the Kleinman-Bylander form used here.

In Fig.~\ref{fig2}, we display  the values of
the high-frequency-dielectric constants of silicon and gallium arsenide
computed with our method at ${\bf E} = {\bf 0}$  for different sizes of the
LRs.
Fitting the data with an exponential function
$\epsilon^\infty=A \exp(- (a_{LR}/a)/\alpha)+B$, we obtain
in the limit $a_{LR} \rightarrow \infty$
the values $B = 13.4 \pm 0.2$ for
silicon and $B = 11.6 \pm 0.1$ for gallium arsenide. The LRT results are
$12.9$ and $11.4$ respectively, thus, supporting the use of
LWFs in the study of the dielectric
properties of solids.
The convergence of the dielectric constant with increasing values of $a_{LR}$
is slower than that
of structural properties~\cite{fdmb}.
The exponential fitting curves correspond to the parameters:
$A = -13.2 \pm 0.3$, $\alpha = 0.81 \pm 0.05$ for
silicon and $A = -10.4 \pm 0.2$, $\alpha = 0.83 \pm 0.05$  for gallium
arsenide.
For the largest localization size
considered ($a_{LR}/a = 2$ which contains $512$ atoms) we obtain
$\epsilon^\infty = 12.3$ for
silicon and $\epsilon^\infty = 10.6$ for gallium arsenide,
which underestimate the LRT results by $5 \%$ and $ 7 \%$ respectively.

In Fig.~\ref{fig3}(a), we show the polarized LWF computed at the theoretical
lattice constant, $a_{LR}/a = 34/24$, and with an electric
field $e \left| {\bf E} \right| = 10^{-3}$ Ha/a.u.
in the $[1 0 0 ]$ direction for silicon and gallium arsenide.
The electric field displaces the center of each LWF  with
respect to the zero-field case, as indicated
with arrows in Fig.~\ref{fig3}(b).
By summing over the four functions within the unit cell at the origin,
the individual displacements
perpendicular to the field cancel out and the induced macroscopic
polarization is parallel
to the field, as expected for solids with cubic symmetry.

In conclusion we have shown that LWFs, a key ingredient of
several `order N' methods for electronic structure calculations,
also allow one to study solids in presence of a finite electric field.
The induced charge density and the dielectric constant obtained from
finite differences converge, in the limit of large LRs and
small electric fields, towards the LRT values.
Furthermore a first example of approximate, {\it polarized}
WFs has been provided.

This work was supported by the Swiss National Science Foundation under
grant No. 20-39528.93 and No. 21-49486.96. We are very grateful to
Massimiliano Di Ventra and Francesco Mauri for many stimulating discussions.
The calculations have been performed on the NEC-SX4 of the Swiss Center
for Scientific Computing (CSCS) in Manno.


%
\begin{figure}[\protect{t}]
\caption[]{
Valence charge density $\Delta n({\bf r})$ (in units of e/cell) induced
by a macroscopic
electric field ${\bf E}$ in the $[100]$ direction (with $e \left| {\bf E}
\right| = 10^{-3}$ a.u.).
Panel (a): Comparison of results obtained
for two different values of $a_{LR}/a$ (the number of atoms inside a LR is
in square  brackets) with the LRT result. The induced
density is represented along the $[111]$ direction at the theoretical
unit cell volume and $d = \sqrt{3} a$.
Panel (b): Contour-plots in the $(0 \bar{1} 1)$ plane of $\Delta n({\bf r})$
for the largest LR size.
The contours are drawn at constant intervals of $0.02$ e/cell.
The thicker contour denotes $\Delta n({\bf r}) = 0$, the continuous (dashed)
contours are for $\Delta n({\bf r}) > 0$ ($\Delta n({\bf r}) < 0$).
\label{fig1}}
\end{figure}

\begin{figure}[\protect{t}]
\caption[]{
Linear dielectric constant $\epsilon^\infty$ of
silicon and gallium arsenide computed with LWFs using different LR
sizes. The values of $\epsilon^\infty$ for $a_{LR} \rightarrow \infty$
are obtained by numerical extrapolation.
$\epsilon^\infty_{LRT}$
indicates the LRT value.
Calculations have been performed at the theoretical lattice
constants~\protect{\onlinecite{fdmb}}: $a_0=10.20$ a.u. for silicon and
$a_0=10.48$ a.u. for gallium arsenide.
\label{fig2}}
\end{figure}

\begin{figure}[\protect{t}]
\caption[]{
Panel (a): Contour-plots in the $(0\bar{1}1)$ plane of the polarized LWFs
in a.u. obtained for an electric field $e \left| {\bf E}
\right|= 10^{-3}$ a.u.
oriented along $[1 0 0]$.
These polarized LWFs are constrained to be zero outside a LR with size
$a_{LR}/a= 34/24$ and centered
on the bond oriented along the $[111]$ direction. The contour
intervals are $0.01$ a.u.
for negative values (dashed lines) and $0.1$ a.u. for positive
values (continuous lines).
Panel (b): Contour-plots of the variation of the
LWFs $\langle {\bf r} |\Delta v_{0,n=1} \rangle = \langle {\bf r} | v^{{\bf
E}}_{0,n=1} \rangle
- \langle {\bf r} | v^{{\bf E}={\bf 0}}_{0,n=1} \rangle$.
The contour interval is $10^{-3}$ a.u. for silicon and
$2 \times 10^{-4}$ a.u. for gallium arsenide.
The LWF centers are indicated
by empty circles while their
variation is depicted by arrows whose
length has been magnified by a factor $100$.
\label{fig3}}
\end{figure}


\begin{thebibliography}{99}

\bibitem{mo}
R. M. Martin and G. Ortiz,
Phys. Rev. B {\bf 56}, 1124 (1997).



\bibitem{LRT1} S. Baroni, P. Giannozzi, and A. Testa, Phys. Rev. Lett.
{\bf 58}, 1861 (1987);  P. Giannozzi, S. de Gironcoli, P. Pavone, and S.
Baroni,
Phys.\ Rev.\ B {\bf 43}, 7231 (1991).

\bibitem{LRT2} S. de Gironcoli, S. Baroni, and R. Resta, Phys. Rev. Lett.
{\bf 62}, 2853 (1989).

\bibitem{Dalcorso} A. Dal Corso and F. Mauri,
Phys.\ Rev.\ B {\bf 50}, 5756 (1994); A. Dal Corso, F. Mauri, and A. Rubio,
{\it ibid.} {\bf 53}, 15638 (1996).

\bibitem{Gonze} X. Gonze, Phys. Rev. B {\bf 55}, 10337 (1997).

\bibitem{Kaxiras}
D. Kandel and E. Kaxiras, Phys. Rev. Lett. {\bf 76}, 1114 (1996).

\bibitem{diref} K. Kunc and R. Resta,
Phys. Rev. Lett. {\bf 51}, 686 (1983);
R. Resta and K. Kunc,
Phys. Rev. B {\bf 34}, 7146 (1986).

\bibitem{bernardini} F. Bernardini, V. Fiorentini, and D. Vanderbilt,
Phys. Rev. Lett. {\bf 79}, 3958 (1997).

\bibitem{Nunes}
R. W. Nunes and D. Vanderbilt,
Phys. Rev. Lett. {\bf 73}, 712 (1994).

\bibitem{nenciu}
G. Nenciu, Rev. Mod. Phys. {\bf 63}, 91 (1991).

\bibitem{mauri}
F. Mauri, G. Galli, and R. Car,
Phys.\ Rev.\ B {\bf 47}, 9973 (1993), and references therein.

\bibitem{fdmb}
P. Fern\'andez, A. Dal Corso, A. Baldereschi, and F. Mauri,
Phys. Rev. B {\bf 55}, R 1909, (1997).

\bibitem{ggg}
X. Gonze, Ph. Ghosez, and R. W. Godby,
Phys.\ Rev.\ Lett.\ {\bf 74}, 4035 (1995).

\bibitem{hk}
P. Hohenberg and W. Kohn,
Phys.\ Rev.\ {\bf 136}, B864 (1964).


\bibitem{notaE} ${\bf E}$
indicates the macroscopic
component of the {\em total internal} electric field.


\bibitem{marzari} N. Marzari and D. Vanderbilt,
Phys. Rev. B {\bf 56}, 12 847 (1997).

\bibitem{pol} R. D. King-Smith and D. Vanderbilt,
Phys. Rev. B {\bf 47}, 1651 (1993);
R. Resta, Rev. Mod. Phys. {\bf 66}, 899 (1994).


\bibitem{ldanota}
We used the
exchange-correlation functional of Ceperley-Alder
as parametrized in J. Perdew and A. Zunger,
Phys. Rev. B {\bf 23}, 5048 (1981).

\bibitem{kb}
L. Kleinman and D. M. Bylander,
Phys. Rev. Lett. {\bf 48}, 1425 (1982).

\bibitem{nota2} We used a kinetic energy cut-off of $24$ Ry to
expand the wave-functions in a plane-wave basis and a 28 ${\bf k}$-point
sampling of the irreducible Brillouin zone to compute the first-order
variation of the valence charge density due to the electric field.

\end{thebibliography}
\end{document}